\documentclass[letterpaper,twocolumn,10pt]{article}
\usepackage{usenix2019_v3}

\usepackage{tikz}
\usepackage{amsmath}

\usepackage{cite}
\usepackage{graphicx}
\usepackage{textcomp}
\usepackage{bmpsize}
\usepackage{xcolor}
\usepackage[utf8]{inputenc}
\usepackage[T1]{fontenc}
\usepackage{soulutf8}
\usepackage{hyperref}
\usepackage{verbatim}
\usepackage{enumitem}
\usepackage{xspace}

\usepackage{tabularx,booktabs}
\newcolumntype{C}{>{\centering\arraybackslash}X} 

\newcommand{\name}{\textsf{SERENIoT}\xspace}

\begin{document}

\date{}

\title{\Large \bf SERENIoT: Collaborative Network Security Policy\\ Management and Enforcement for Smart Homes}

\author{
{\rm Corentin Thomasset}\\
Polytechnique Montreal\\Montreal, Canada \\corentin.thomasset@polymtl.ca
\and
{\rm David Barrera}\\
Carleton University\\Ottawa, Canada \\david.barrera@carleton.ca
} 

\maketitle

\begin{abstract}
Network traffic whitelisting has emerged as a dominant approach for securing consumer IoT devices. However, determining what the whitelisted behavior of an IoT device \textit{should be} remains an open challenge. Proposals to date have relied on manufacturers and trusted parties to provide whitelists, but these proposals require manufacturer involvement or placing trust in an additional stakeholder. Alternatively, locally monitoring devices can allow building whitelists of observed behavior, but devices may not exhaust their functionality set during the observation period, or the behavior may change following a software update which requires re-training. This paper proposes a blockchain-based system for determining whether an IoT device is behaving like other devices of the same type. Our system (\name) overcomes the challenge of initially determining the correct behavior for a device. Nodes in the \name public blockchain submit summaries of the network behavior observed for connected IoT devices and build whitelists of behavior observed by the majority of nodes. Changes in behavior through software updates are automatically whitelisted once the update is broadly deployed. Through a proof-of-concept implementation of \name on a small Raspberry Pi IoT network and a large-scale Amazon EC2 simulation, we evaluate the security, scalability, and performance of our system.
\end{abstract}

\section{Introduction}
    \label{sec:introduction}
    The rapid adoption of the Internet of Things (IoT) challenges well-established computer security strategies. Because of their deployment scale, IoT devices cannot be secured using traditional techniques such as anti-malware or network intrusion detection systems (NIDS). The diversity in IoT hardware, software, and network footprint, combined with the deployment volume makes it difficult to design security systems that are effective yet not overburdened with management complexity. This is of particular importance in smart homes, where users are typically not security experts.   
    
    IoT devices are pervasive~\cite{kumar_all_nodate} and always connected. They are manufactured to be low-cost, so security is often not the primary design goal. As expected, numerous papers~\cite{alrawi_sok:_2019, fernandes_security_2016, noauthor_hack_nodate, noauthor_elaborate_nodate} studying the security of IoT devices have found a steady stream of vulnerabilities (e.g., the Mirai botnet~\cite{antonakakis_understanding_nodate}) that pose a threat to users, to their environments, and to the broader global Internet infrastructure. 
    
    While IoT devices are diverse, one unifying characteristic is that their feature-set is generally simple; a device may sense its environment and submit readings to a cloud service (e.g., a humidity sensor), listen for inbound requests to perform some action (e.g., a WiFi light switch), or some combination of both. IoT devices by definition are not general purpose computers\footnote{We note, however, that some IoT devices may be built on top of general purpose operating systems such as Linux.}, and as such they do not require the network privileges of a general purpose system to perform their primary task. However, IoT devices are often treated indifferently from mobile phones, laptops, and other general purpose systems on networks, allowing any network communication that originates from the device to reach any host on the Internet. This over privilege allows compromised devices to directly attack remote hosts and services, or to act as steppingstones in more sophisticated attacks.
    
    To prevent these attacks, existing NIDS systems could be used, but these require the user to configure operating parameters and tune the detection logic to avoid being overwhelmed by false positives. A conceptually simpler approach is the idea of allowing only a small set of network traffic to flow to/from an IoT device as needed. By \textit{whitelisting} types of network activity that a device can generate or accept (which should roughly match the functional requirements of the device's primary task), the device can be constrained without requiring the modification of its software. This is of particular interest in IoT, where devices may have long lifespans sometimes outlasting the manufacturer or software update support period. Moreover, false positives (i.e., blocking legitimate outbound connections) should be few and far between if the device can be accurately profiled. 
    
    Manufacturer Usage Descriptions (MUD\cite{lear_manufacturer_nodate}) standardize the policy language in which IoT whitelists can be written, so that the device manufacturer or a trusted third party can encode device behavior into a machine-readable policy. This policy can be enforced at the network edge, protecting all devices in the local network. The open question that remains is: \textit{what network behavior should be included in the whitelist?} Asking manufacturers to provide whitelists may not scale; there are too many unique IoT vendors, some of which simply re-brand devices manufactured by another vendor. A trusted third party could analyze devices and generate whitelists, but the business incentives (including user willingness to pay for such a service) aren't clear. Users themselves could analyze local device behavior and generate profiles, but this approach may not scale to households with large number of IoT devices.
    
    In this paper we propose a blockchain-based network security policy management and enforcement system for home IoT environments. Our system, \name (pronounced \textit{Serenity}), characterizes IoT device behavior locally and uses a decentralized ledger to determine whether the local behavior matches that observed by other peers in the network. Policies (whitelists of allowed behavior) are the result of a consensus algorithm identifying network behavior observed by the majority of nodes in the network. Network connections that are unique to a device are blocked until they are observed by most nodes, preventing the spread of Mirai-style botnets. 

    \name is designed to run on network appliances such as home routers. The system analyses IP traffic between the local IoT devices and their cloud APIs, making it compatible with all IP IoT devices and hubs. As these appliances are already present in home networks (e.g., ISP-provided home routers), our system doesn't require any drastic network topology change and integrates directly into actual homes. It is designed to extend the security features of home gateways by adding network policy enforcement for IoT devices. \name takes advantage of the IoT devices having limited sets of actions and being widely spread. It leverages the blockchain to build a decentralized ledger of trusted connections and blocks all connections that are not explicitly in the whitelist.
    
    Our contributions are:
    \begin{enumerate}
        \item The design and implementation of a novel approach to build a whitelist of behavior of an IoT device. The approach is based on blockchain and requires no opt in by manufacturers or trust in third parties. \name's public blockchain provides new data sources to audit IoT device behaviors at scale and assists in the detection of new threats. 
        \item Evaluation of our system on large-scale simulations with 53 devices and 1000 Sentinels and on a small-scale test bed with real world devices.
    \end{enumerate}
    
    The remainder of the paper is structured as follows. Section~\ref{sec:background} reviews IoT security background and related work and gives a brief overview of relevant blockchain concepts. Section~\ref{sec:serenity} presents the technical details of \name. Section \ref{sec:evaluation} evaluates scalability, performance and security. Sections~\ref{sec:limitations} and \ref{sec:discussion} present the limitations of our implementation and discuss related deployment issues. We conclude in Section \ref{sec:conclusion}.

\section{Related work and Blockchain Review}
    \label{sec:background}
    \subsection{IoT security}
        \label{sec:background:iot_security}
        One common solution to protect IoT networks is to deploy a signature-based Network Intrusion Detection System (NIDS)~\cite{kumar_early_2020, habibi_heimdall:_2017,meidan_n-baiotnetwork-based_2018}  on IoT networks. NIDS monitor network traffic and look for known attack signatures. These solutions are therefore only efficient if the attack is already known and require constant updates to have the latest signature base. 
        Although these solutions might be workable for industrial IoT networks with dedicated security teams, complex IDS solutions are not suited for home environment where experts are likely unavailable to monitor, maintain, and configure them.
        IDSes can be augmented by using machine learning to detect previously unseen attacks. However, this introduces uncertainty as false positives can be exploited by attackers~\cite{sommer_outside_2010}. The accuracy issue is also present when identifying device types~\cite{miettinen_iot_2017}. The similarity in behavior of distinct devices makes it difficult to determine which device generated the traffic, or what policy to apply to a particular device. 
        
        An alternative approach is to whitelist device behavior based on policies describing the devices' expected behavior. This approach is sometimes referred to as specification-based intrusion detection, where the policy is a narrowly defined list of allowed behavior. The policies can be provided by the manufacturers or trusted parties as proposed by the IETF in RFC8520~\cite{lear_manufacturer_nodate} or generated by local device observation~\cite{barrera_idiot:_2017}. Other approaches are to divide devices into controllers and non-controllers, preventing non-controllers (e.g., lightbulbs) connecting from other devices and limiting them to only connect to their cloud endpoints~\cite{goutham_hestia_2019}.
            
        Anomaly detection capabilities can also be embedded into devices themselves~\cite{raza_svelte:_2013}. The idea here is that the firmware on the device is updated to include an anomaly detection agent which monitors the system for malicious activity. Since this approach requires changes to software running on every IoT device, it is largely incompatible with devices that are deployed and no longer maintained. Moreover, it requires strong cooperation with manufacturers for adoption. 
        
    \subsection{Blockchain review}
        \label{sec:background:blockchain}
        We briefly review the key concepts of blockchain technology. A deeper treatment can be found in~\cite{ruoti_sok:_2019}. 
        Blockchain technology addresses use cases where multiple distrusting parties want to jointly participate in a system. Blockchain provides shared governance where participants collaboratively decide what gets added to the chain and ensures that the protocol is being followed correctly by all the participants. Participation may be open (anyone can join, possibly without registration) or closed (only authorized participants can contribute).
        
        A major aspect of blockchains is their verifiable sate: the data in a blockchain reflects the output of its consensus protocol which has been verified by all the participants. That is, only data that has been agreed upon through consensus can be added to the chain, leading the chain to contain only verifiable data. Once data has been verified by participants in the network, a new block containing this data is added to the chain. This data includes a cryptographic link to the previous block, allowing all parties to verify the continuity of the chain. 
        
        The consensus algorithm is thus a key aspect of every blockchain. It ensures that the chain of blocks containing the data is kept synchronized between participants so that they all have an identical copy of it at any time. It also prevents the blockchain to grow almost instantly by introducing a delay between the creation of new blocks. Multiple consensus algorithms exist~\cite{cachin_blockchain_2017}. The proof of work (PoW)~\cite{back_partial_1997, juels_client_1999} algorithm is widely used by popular permisionless (open) blockchains such as Bitcoin\footnote{https://bitcoin.org/} and Ethereum\footnote{https://ethereum.org/} and requires block hashes to be smaller than a defined target. The weight of each participant in the validation process is thus determined by its capacity to compute hashes and this mechanism ensures that participants are randomly selected to create new blocks. However, this approach is very costly, energy and computationally speaking. Indeed, all the effort has no other utility than randomly delaying participants capacity to produce valid blocks. Another approach is the proof of stake~\cite{Kiayias_ouroboros_2017}. Proof of stake doesn't rely on computing hashes and thus, unlike proof of work, doesn't have massive energy requirements. With this algorithm, the creator of a new block is chosen within a pool of participants who have staked a certain amount of cryptocurrencies. The penalty to harm the network is then the cost of losing the staked amount of cryptocurrencies. For major blockchains this can amount to tens of thousands of dollars. A participant trying to take over the network would also need to own 51\% of the cryptocurrency supply on that blockchain. That amount to billions of dollars for major cryptocurrencies. It is thus less likely to happen than controlling half of the network hash power for proof of work~\cite{matonis_bitcoin_2014}. However, this consensus mechanism requires a built-in cryptocurrency to work. Both of these consensus algorithms are widely used and provide a probabilistic way to verify blocks' validity.
        
        Finally, another feature of blockchains is data loss prevention: Thanks to the decentralized nature of blockchains, the data in the chain is replicated across participants which allows recovery in case of data loss. At any time a participant can ask for a copy of the full chain and verify its contents.
        
        Through these properties, Blockchain provides a tamper-proof decentralized ledger that can be used beyond crypto-currencies in applications requiring accountability, transparency and trust in data~\cite{ruoti_sok:_2019}. Blockchain is also suited for open network applications where untrusted participants need to reach a consensus as is the case in collaborative characterization systems.
        Collaborative characterization consists of aggregating multiple observations of similar systems to identify their shared traits. When working with a large set of observations from different sources, it allows to identify the intended behavior of the systems without any prior knowledge. As this behavior is identified using observations on a large population, it is highly resistant to outliers and only reflects the true nature of the system.
        When applied to IoT, collaborative characterization enables us to identify the intended behavior of a device model, represented by the set of packet signatures that are common to the majority of observed IoT devices.

    \subsection{IoT security and blockchain}
        \label{sec:background:iot_blockchain}
        Golomb et al.~\cite{golomb_ciota:_2018} propose CIOTA as a system monitoring devices' behavior at the firmware level to build a ledger of legitimate actions. This is achieved by continuously gathering data from devices using Extensive Markov Models (EMM) and building a collaborative model of device behavior. The Blockchain's ledger is then used to inform a client-side intrusion detection system. While a preliminary security evaluation of CIoTA appears promising, it requires modification of the devices firmware to embed the software EMM agent. \name learns device behavior at the network layer see Section~\ref{sec:serenity:sentinel-architecture}, and thus does not require any changes to the firmware enabling greater compatibility with existing devices.
        
        Mendez Mena et al.~\cite{mendez_mena_blockchain-based_2018} have built and evaluated a proof of concept based on the Ethereum blockchain to protect the edge of home networks. Their "gatekeeper" enforces a whitelist of allowed actions which is computed based on the information stored in an Ethereum smart contract. The smart contract provides a tamper proof ledger to report the local network information gathered by gatekeepers. This approach has several shortcomings: Despite being based on blockchain, it does not solve the problem of policy creation addressed herein. Their approach is also centralized as gatekeepers interact with a single smart contract which could be replaced by a database. Moreover, this solution is not tailored for real world use as its deployment on the Ethereum public blockchain would require payment to push policy updates to the smart contract.

\section{\name Intrusion Detection System}
    \label{sec:serenity}
    \subsection{Overview}
        \label{sec:serenity:overview}
        \name is a collaborative specification-based intrusion detection system for home IoT networks. It monitors the network traffic to/from IoT devices to detect and block anomalous packets and connections. It relies on a decentralized ledger that characterizes devices’ behavior and hosts a list of allowed packet signatures.
        
\begin{figure}[h!]
            \includegraphics[width=\linewidth]{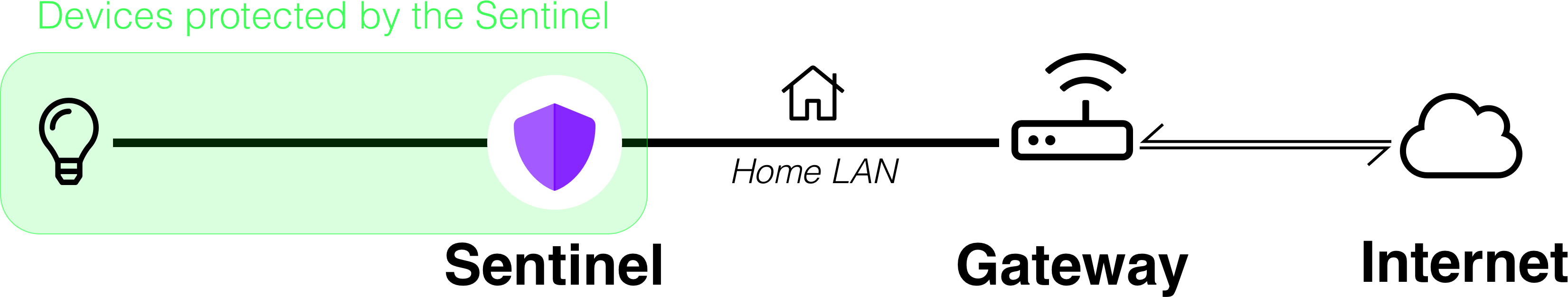}
            \caption{\name Network topology. Sentinels acts as middleboxes between IoT devices and the network gateway, enabling blocking of connections that are outside the device specification.}
            \label{fig:network_topology}
        \end{figure}

        \name nodes (called \emph{Sentinels}) are designed to be deployed on network appliances or middleboxes such as routers. A typical set-up would see one Sentinel deployed per smart home (see Figure~\ref{fig:network_topology}), collaborating with other remote Sentinels to determine the correct network behavior of IoT devices. Sentinels advertise a WiFi network to which IoT devices connect, thus allowing mediation and filtering of all network connectivity between the devices and the Internet. The wireless network operates as a network bridge to the home local area network (LAN), so all traffic entering or leaving the Sentinel is monitored. Through its use of a distributed ledger and peer-to-peer communication, \name can operate with little-to-no user input. Moreover, compared to other network security solutions such as signature-based intrusion detection systems, \name's Sentinels are implicitly always up to date.
        
        Concretely, Sentinels only forward packets that are defined in a \emph{whitelist}. Any network connection that is not specified in the whitelist is discarded by the Sentinel. The whitelist specifies network packet signatures characterizing the behavior of a specific IoT device as observed by the majority of Sentinels on the network. Whitelists for all IoT devices are stored in \name's blockchain. Through the use of blockchain, \name is fully decentralized and can be bootstrapped with a small number of Sentinels. It allows the system to be fully independent from trusted third parties, device manufacturers, and to support a large set of diverse IoT devices. We discuss why we use blockchain in Section~\ref{sec:serenity:blockchain}. \name is fully backward compatible with many existing IP-based IoT devices, requiring no changes to their hardware, firmware, or apps. The system is also designed to be forward compatible with devices that don't yet exist, as long as they use IP-based communication and can connect to the local Sentinel. 
        
\begin{figure}[h!]
                \includegraphics[width=\linewidth]{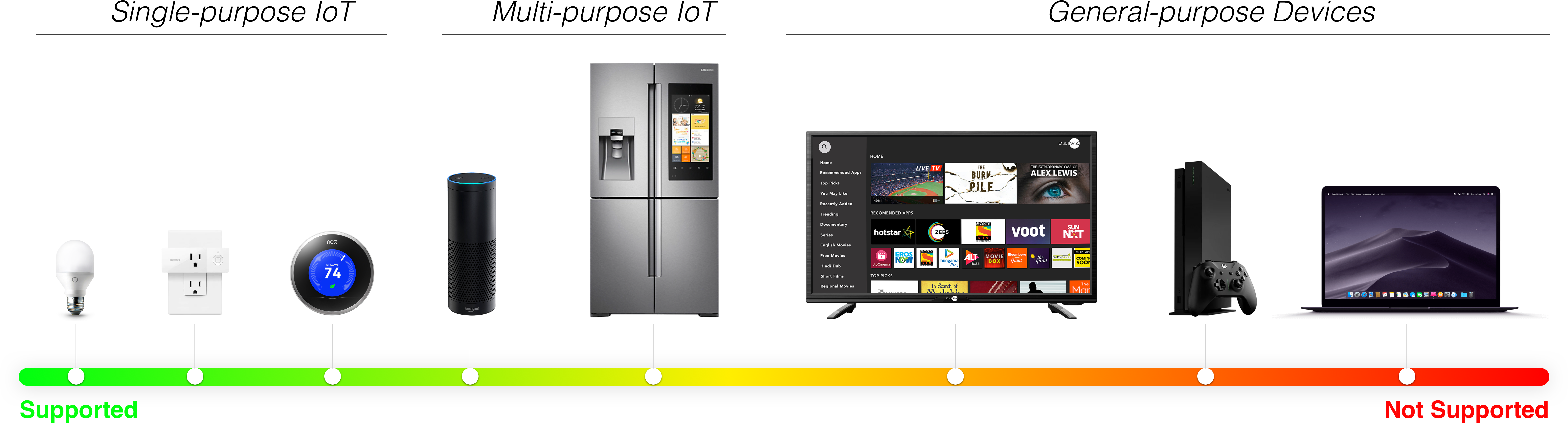}
                \caption{Supported devices for \name. Devices toward the left have simpler network behavior and tend to have a similar network footprint shared across devices of the same type. Devices toward the right have unique network footprints determined by their users.}
                \label{fig:devices}
            \end{figure}

        We built \name with a focus on consumer IoT, noting high accuracy and performance when working with feature-limited devices such as smart bulbs, smart switches, smart locks, smart thermostats, etc. (see Figure ~\ref{fig:devices}). These devices typically only interact with a small set of cloud services through well-defined APIs, thus their network footprint can be accurately determined (see Section \ref{sec:evaluation:virt}). According to a 2019 study~\cite{kumar_all_nodate}, this should cover approximately 41\% of devices deployed in North American homes, and 28.4\% of devices in Western Europe\footnote{We include all IoT devices in the study by Kumar et al.~\cite{kumar_all_nodate} except media boxes, game consoles, and file storage appliances which are functionally as complex as general purpose computers}. \name cannot support systems with variable (typically human-dependent) network behavior, since each system may create a unique set of network connections. 
        
    \subsection{Threat model}
        \label{sec:serenity:threat-model}
        \name protects devices against attackers trying to change their behavior, as widely used by botnets~\cite{antonakakis_understanding_nodate, jun_29_your_2018}. \name has been designed to counter the two following attack scenarios:
        \begin{itemize}
            \item An IoT device has been compromised locally by a malware trying to change its behavior to accomplish malicious actions. The attack vector can vary;  the IoT device can be infected by another device on the local network (for example by an infected computer or IoT device), or the infection can be the result of a physical action on the device (for example a memory card swap). In this situation \name would protect the IoT device from attacking targets on the internet by blocking all the outgoing traffic deviating from the specification.
            \item An IoT device is directly exposed on the internet. \name would protect the IoT device from incoming attacks by blocking all incoming traffic differing from the specification. Most IoT devices don't normally receive incoming connections from the internet and \name will then behave as a firewall blocking all incoming connections. 
        \end{itemize}

        Our prototype does not currently monitor local network traffic and does not protect against malicious usage (for example an attacker using the user credentials to view a smart camera video feed) as this kind of attack meets the specifications of allowed actions.
        
        Once a whitelist has been populated for a device, an attacker would need to change the behavior of more than 50\% of the IoT devices of the same type to change the specifications and allow the attack to go through (see Section~\ref{sec:evaluation:security}).

    \subsection{Sentinel architecture}
        \label{sec:serenity:sentinel-architecture}
        \begin{figure}[h!]
            \includegraphics[width=\linewidth]{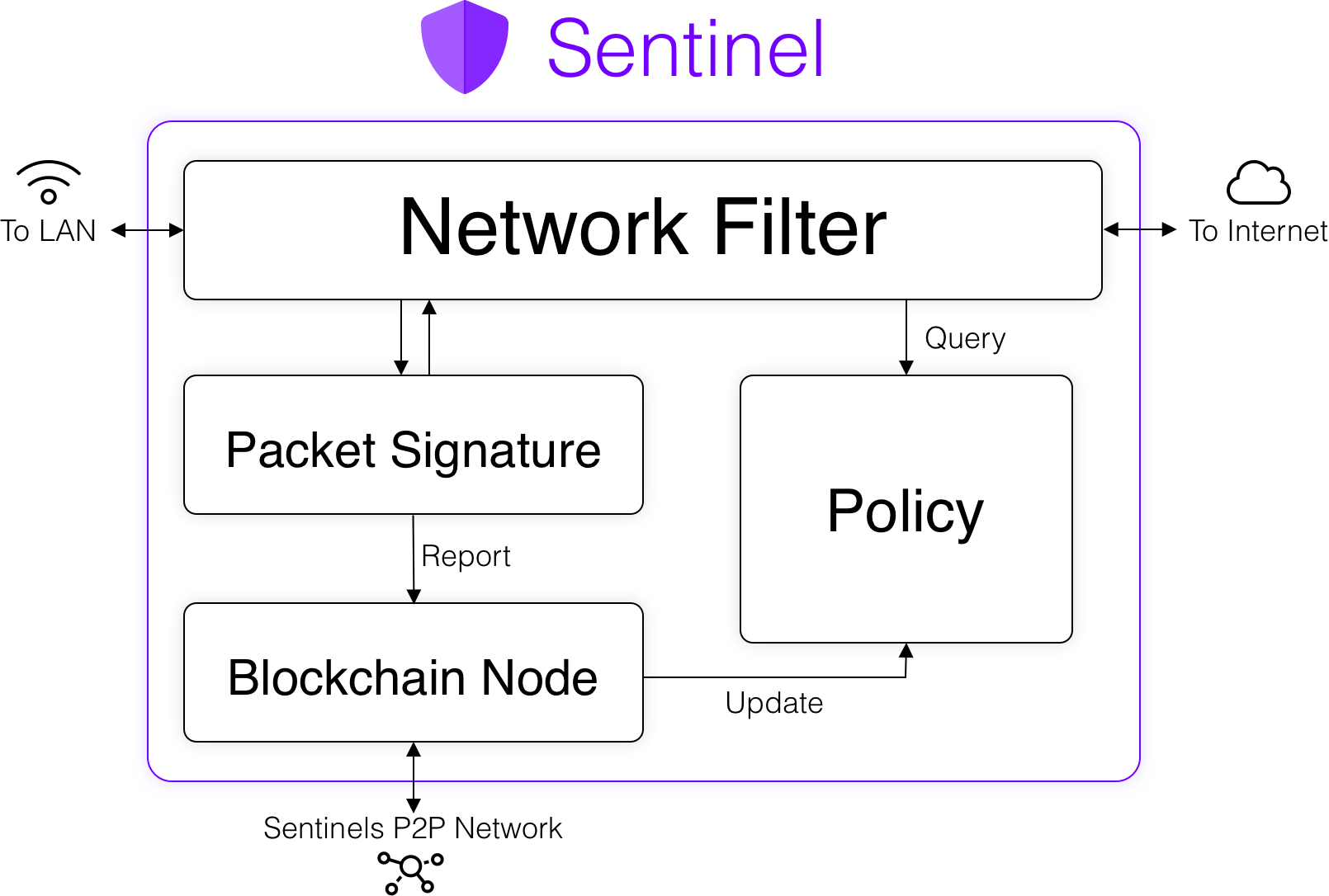}
            \caption{Main components of a Sentinel. Description inline.}
            \label{fig:overview}
        \end{figure}

        Sentinels use a modular architecture with 4 main components (see Figure~\ref{fig:overview}): The network filter component (1) is in charge of enforcing the policy by dropping network packets that are not allowed and forwarding acceptable traffic. The network filter relies on the packet signature module (2) to serialize the raw IP packet into a textual signature and on the policy module (3) that lists all the allowed packet signatures. Finally, the blockchain module (4) keeps the policy updated by synchronizing the ledger with the other Sentinels and by reporting the newly recorded packet signatures. 
            
    \subsection{Computing packet signatures}
        \label{sec:serenity:signatures}
        Packet signatures allow Sentinels to characterize recorded packets. An effective signature algorithm should be precise enough to differentiate packets from different network connections but flexible enough to produce the same signature across devices of the same model. Unlike general purpose computers whose behavior changes depending on their usage, IoT devices of a same model behave similarly (often identically in terms of traffic transmitted and destination) and produce similar network traces. We have verified this hypothesis during our evaluation in Section~\ref{sec:evaluation}.
        Our proof of concept uses packet signatures at the IP level and focuses on the fields that remain constant across devices. While other packet signatures and connection fingerprinting techniques exist, we use a NetFlow\footnote{https://www.cisco.com/c/en/us/products/collateral/ios-nx-os-software/ios-netflow/prod\_white\_paper0900aecd80406232.html} like representation to provide a balance between uniqueness and consistency across devices. \name's packet signature aggregates sequences of packets sharing the following values: 
        \begin{itemize}
            \item Protocol of the IP payload
            \vspace{-0.2cm}\item Endpoint (domain name or IP address if domain is unavailable)
            \vspace{-0.2cm}\item Service port
        \end{itemize}
        
        The Endpoint identifies the remote host with which the IoT device is interacting. To resolve potential domains, we perform reverse DNS lookups. The service port identifies the well-known port number used by the connection\footnote{The service port generally refers to the remote endpoint's port. However, some IoT devices (e.g., cameras) host certain services locally, in which case the service port refers to the local port hosting the service. To differentiate between local and remote services, we append a direction identifier (L for local or R for remote) to the service port.}. Packets of a same flow will share the same signature that will be used by \name to identify anomalous flows and packets. Signatures are computed by serializing and hashing the following values: 
        \[Signature = H(protocol, endpoint, service~port)\]
        
        Packet signatures don’t include device-specific identifiers such as Media Access Control Organizational Unique Identifiers (OUI). Indeed, it is unclear whether all devices of a same model will share a single OUI as many manufacturers are allocated more than one. Two identical devices with different OUIs would be assigned to different chains weakening the security of both chains.  Moreover, malicious code running on an IoT device may be capable of manipulating the MAC address.
        Our choice of packet signatures allows to differentiate packets going to untrusted hosts from those going to the manufacturer's API. It also allows to differentiate packets initiated by the monitored device from those initiated by a remote entity in the case of IPv6 network or networks without NAT where devices are directly exposed on the internet.
        
        Note that \name doesn't precisely identify devices. Devices that produce the same set of packet signatures are grouped and the system assumes they are of the same type. Devices are characterized by their packets signatures and device types are fingerprinted by hashing their sorted set of packet signatures.

    \subsection{\name's blockchain}
        \label{sec:serenity:blockchain}
        The blockchain is the key behind \name's collaborative policy generation mechanism. It ensures that all the packet signatures written to the whitelist are agreed upon through a distributed consensus protocol. This provides robustness and trust by making sure malicious signatures do not become whitelisted as long as a majority of Sentinels participating in the network observe legitimate behaviors on their local IoT devices. 
        
        Our choice of building \name on blockchain stems from 3 core design requirements:
        (1) It allows the system to run without relying on trusted third parties such as manufacturers. It is indeed unlikely that a third party will provide a system supporting all past, present and future IoT devices from different brands without expecting the user to pay a subscription fee. (2) Deploying up a highly available infrastructure that provides real time security policy updates to potentially millions of users would be costly, whereas the use of blockchain distributes the system across nodes. (3) The system should allow public open access and should not be restricted to users of specific vendors. We evaluate the scalability and robustness of our blockchain-based system in Section~\ref{sec:evaluation}.

        To implement \name, we have designed a custom public blockchain based on Bitcoin's blockchain principles but with no inherent cryptocurrency. We elected to build a custom blockchain because current blockchain frameworks are either strongly tied to token economics (e.g., Ethereum\footnote{https://ethereum.org/}) or designed to build permissioned blockchains (e.g., Hyperledger fabric\footnote{https://www.hyperledger.org/projects/fabric}). Designing our own chain gives us the flexibility to include only features specific to our requirements while avoiding compatibility challenges arising from trying to retrofit another framework to our use case. 

        \subsubsection{Ledger}
            \label{sec:serenity:blockchain:whitelist}
\begin{figure}[h!]
                \includegraphics[width=\linewidth]{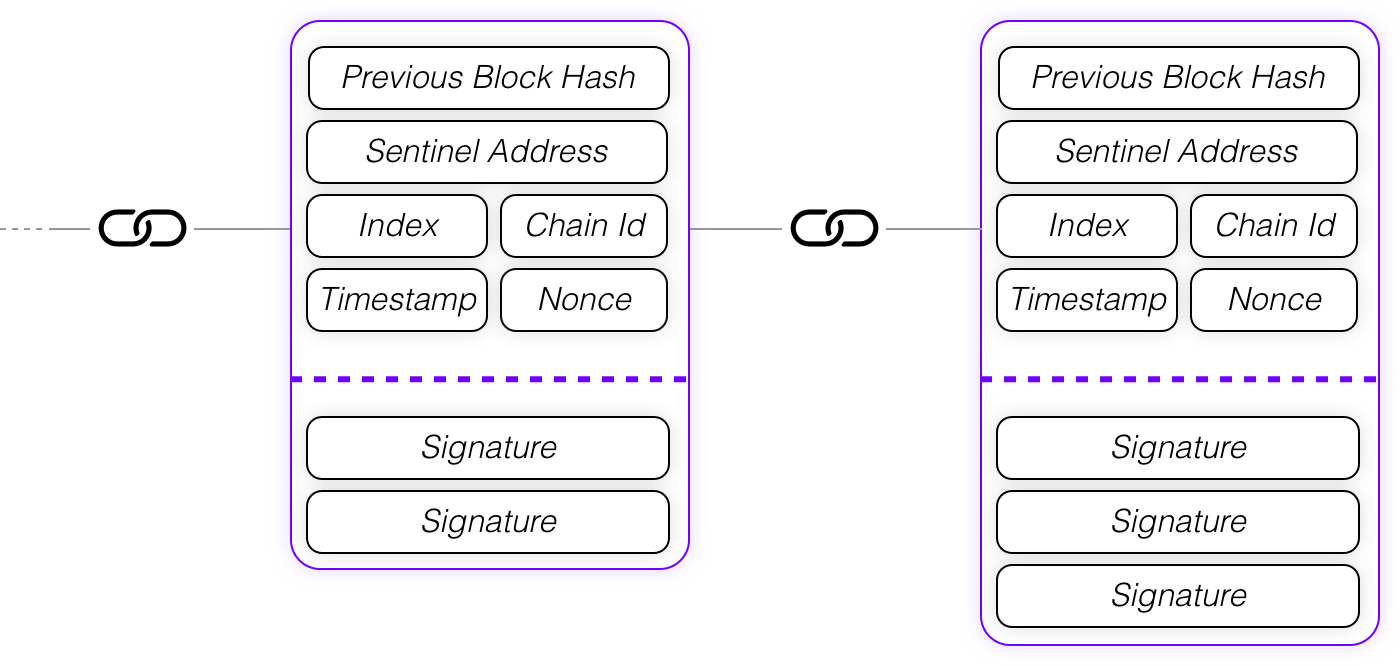}
                \caption{\name's Block architecture. The Sentinel Address is a unique identifier generated at Sentinels start up using the same algorithm used to generate Bitcoin's addresses.}
                \label{fig:blocks}
            \end{figure}

            \name's ledger includes packet signatures reported by the Sentinels. It is based on a distributed timestamp server chaining data blocks together. The linked timestamping mechanism ensures that blocks cannot be rearranged or modified without invalidating subsequent blocks in the chain. As the blockchain grows, Sentinels converge on the chain with the most blocks. In our implementation, blocks store a list of packet signatures reported by the Sentinels instead of a Merkle root of transactions as in Bitcoin (see Figure~\ref{fig:blocks}). The complete list of reported signatures is indeed necessary to build the policy and there is thus no need for selective reveal.
            
            \begin{figure}[h!]
                \includegraphics[width=\linewidth]{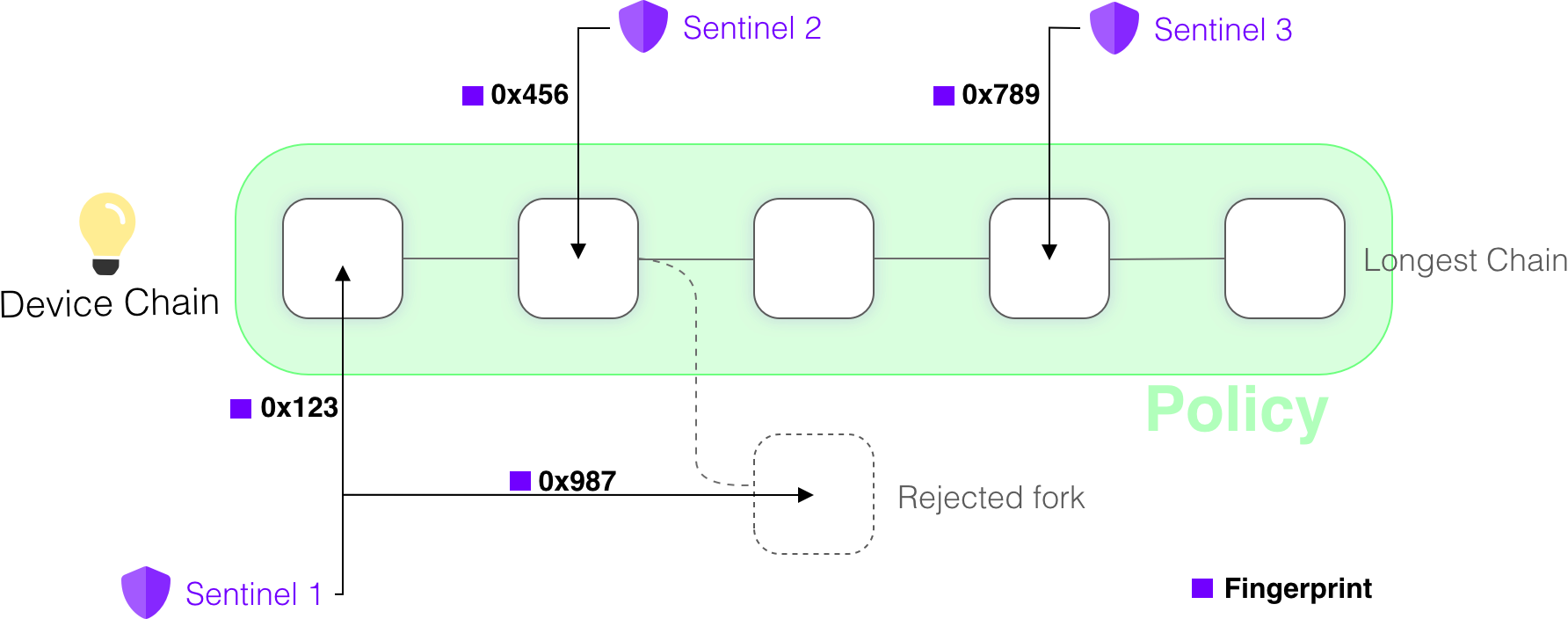}
                \caption{Device chain: The Sentinels add packet signatures into blocks. The chain grows and only signatures listed into the longest chain are trusted.}
                \label{fig:device-chain}
            \end{figure}
        
            To extend the chain and report new packet signatures to the system, Sentinels only work on top of blocks that contain signatures that they have previously observed. That is, Sentinels avoid appending to chains that include unknown signatures. These packet signatures may be malicious or reflect previously unseen connections for a device. Thus, the fastest growing chain always contains the most common packet signatures that have been observed by a majority of Sentinels. This mechanism is described in detail in Section~\ref{sec:serenity:blockchain:sentinel_process}
            
            The whitelist is a cumulative set of allowed packet signatures that have been included in confirmed blocks since the chain genesis. Note that this whitelist may allow behavior that is no longer necessary to the device to operate (e.g. a feature that was removed through a software update). Future work will explore removing outdated signatures.

        \subsubsection{Consensus}
            \label{sec:serenity:blockchain:consensus}
            \name's consensus algorithm ensures that the whitelist is kept synchronized between Sentinels so that they all converge to an identical copy of the blockchain. It is also responsible for making sure that the fastest growing chain gathers the most Sentinels. To facilitate the development of our proof of concept, we implement a proof of work consensus algorithm~\cite{nakamoto_bitcoin:_nodate}. We discuss alternative consensus algorithms in Section~\ref{sec:discussion:consensus}.
            
            With proof of work, blocks are produced by nodes racing to solve computational puzzles. The node that solves the problem appends its block to the chain. Each additional block increases the effort required to rewrite the longest chain, since changing a past block would require every subsequent proof of work to be recomputed. As long as the computational power distribution remains balanced across Sentinels, the fastest growing chain will gather the most Sentinels.

        \subsubsection{Supporting multiple devices}
            \label{sec:serenity:blockchain:mult_devices_support}
            The logic described previously works well for one specific IoT device. Indeed, every IoT device protected by \name needs its own blockchain/whitelist as Sentinels cannot adjudicate on blocks containing packet signatures for unknown devices. \name uses a multichain solution allowing Sentinels to subscribe to the whitelists concerning the devices they protect. Thus, each device type uses a separate blockchain to track its behavior. When a new device is connected to a Sentinel, the Sentinel profiles it and assigns it to a chain aggregating similar devices. To profile a device and subscribe to the right whitelist, Sentinels observe the device behavior during a short profiling phase upon its connection. Once the profiling phase is over, Sentinels compute the whitelist identifier corresponding to the device. The whitelist identifier is computed by hashing the sorted list of packet signatures collected during the profiling phase. \name doesn’t identify precisely devices. Devices that produce the same set of packet signatures are grouped together and the system assumes they are the same type. Thus, devices are characterized by their packets signatures and device models are fingerprinted by hashing their sorted set of packet signatures.
            
            To support multiple IoT devices, \name's Blockchain is composed of one control chain and multiple device-specific chains (also referred to as whitelists), one for each device model protected by the Sentinels. In our implementation, the control chain stores the block headers of valid whitelist blocks and uses the proof of work consensus mechanism. Device specific chains do not have an independent consensus mechanism, they instead leverage the control chain's proof of work.
            
            \begin{center}
                \begin{figure}[h!]
                    \includegraphics[width=7cm]{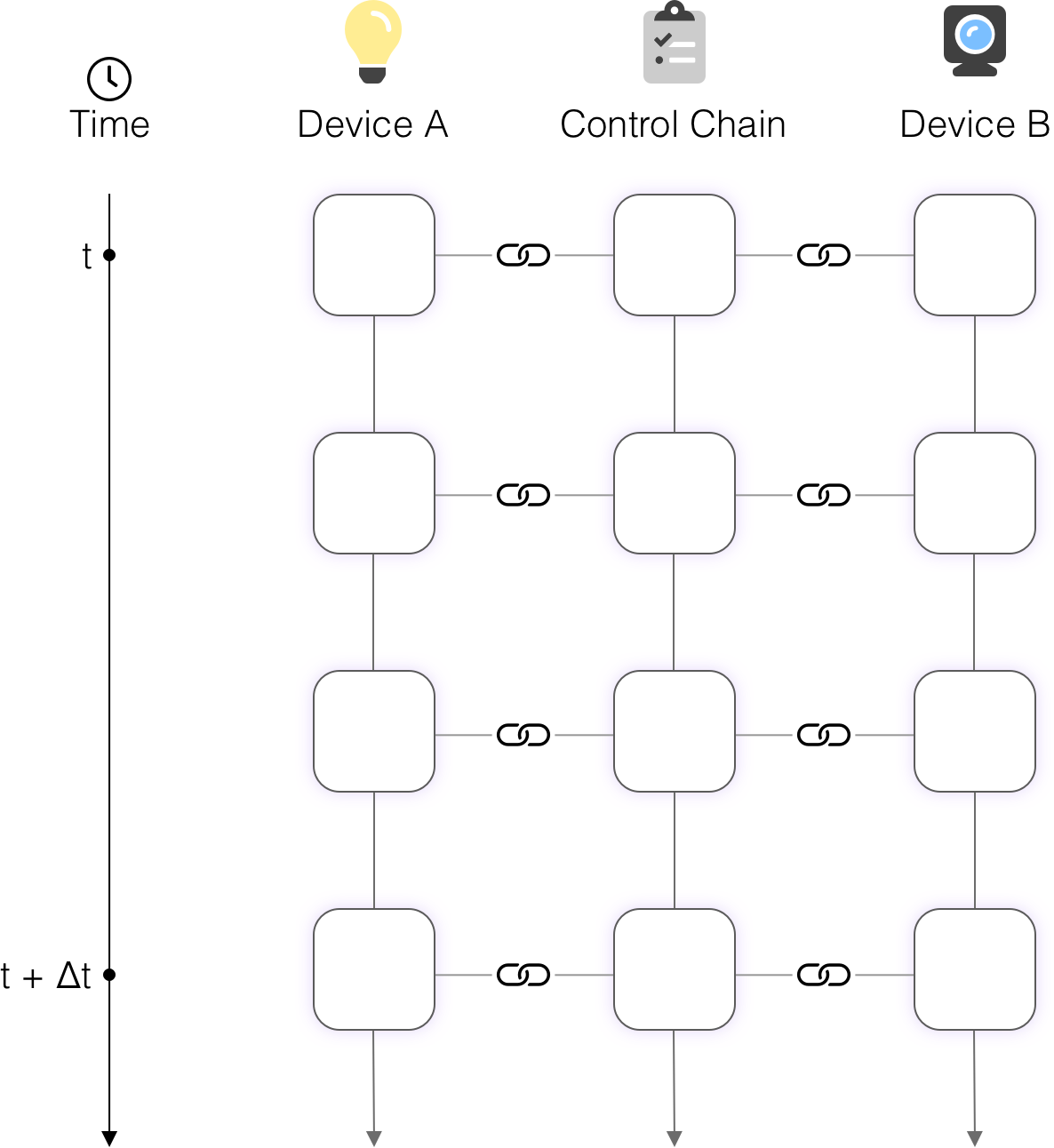}
                    \caption{Multichain support in \name. Block headers of device specific blockchains are incorporated into a single control chain.}
                    \label{fig:multichain}
                \end{figure}
            \end{center}
                
            The control chain improves robustness by requiring all Sentinels (regardless of their locally monitored IoT devices) to ultimately contribute to a single master chain while building on device-specific blockchains. It makes it harder for a malicious actor to target a specific unpopular whitelist to rewrite it. Indeed, all blocks in device-specific whitelists are validated in the control chain that gathers all the Sentinels of the network.
        
        \subsubsection{Sentinels workflow}
            \label{sec:serenity:blockchain:sentinel_process}
            Sentinels participate in maintaining and updating the whitelists by serving as blockchain nodes. Sentinels only subscribe to the whitelists corresponding to the devices they locally monitor. 
            A Sentinel's blockchain node process can be described as follows.
            \begin{enumerate}
                \item The Sentinel monitors IoT devices that are connected to it and collects new packet signatures into whitelist block candidates. One whitelist block candidate is created per subscribed whitelist. If a device is inactive or no new packet signatures have been recorded, the Sentinel builds an empty  block.
                \item The Sentinel computes the hashes of the whitelist block candidates' headers and adds them to a control block candidate.
                \item The Sentinel works on solving the proof of work for the control block candidate.
                \item The first Sentinel to produce a control block is selected to append its whitelist block candidates to the corresponding whitelists. To do so, it broadcasts the control block along with all the whitelist blocks listed within.
                \item Sentinels always accept broadcasted control blocks. Sentinels only accept a broadcasted whitelist block if they are registered to the corresponding whitelist and if they recognize all its packet signatures. When accepting a block, Sentinels work on extending the chain on top of it. A whitelist block is only valid if its block header is listed in a control block. Sentinels always converge on the longest chain and forks are resolved when a branch becomes longer than the others. 
            \end{enumerate}
            
        \subsubsection{Adding incentive for open networks}
            \label{sec:serenity:blockchain:incentive}
            \name's blockchain has been designed to work with no inherent cryptocurrency. Thus, Sentinels contributing to the network by providing computational power cannot be rewarded with some cryptocurrency. To encourage Sentinels to stay active and contribute to the blockchain, inactive Sentinels are isolated by their neighbors and do not receive the latest whitelists updates. To signal their activity and contribution to the network, Sentinels use a mechanism inspired by mining pools and broadcast partial solutions of the proof of work they are trying to solve. This proves to their neighbors that they are active and contributing to the system.

    \subsection{Detecting behavior changes}
        \label{sec:serenity:behavior_change}
        \name is designed to protect IoT devices with limited functionalities such as smart bulbs, smart plugs or smart cameras. These devices typically establish a small set of network connections so we can characterize their expected behavior by observing the network traffic of a large number of devices of a specific model. Sentinels use the most common behavior observed by nodes in the system to create a specification of what the observed device should be allowed to do. Specification-based intrusion detection systems raise alarms when behavior deviates (even slightly) from a narrowly defined specification. Network traffic is either permitted or blocked, with no notion of confidence or likelihood of attack, as is the case with anomaly-based IDS. Specification-based IDS is also different from signature-based IDS, where experts define signatures of all known attacks. \name seeks to define signatures of known good behavior and block all other network traffic. 
        
        To do so, Sentinels use whitelists to file the packet signatures characterizing devices' intended behaviors. When a packet from an IoT device is recorded by a Sentinel, the Sentinel computes its signature and verifies whether the signature exists in the whitelist associated with the device. If the signature is trusted, the packet is forwarded to its destination. Otherwise the packet is blocked, and the Sentinel reports the packet signature (i.e. adds it to current block candidate of the whitelist). It will eventually be appended to the whitelist if the majority of Sentinels also report it.
        
        This mechanism allows Sentinel to detect and block anomalous behaviors that are only observed on a small proportion of the monitored devices.
 
        \subsubsection{Updating the policy}
            \label{sec:serenity:behavior_change:learning}
            When a device' behavior changes, other Sentinels on the network report whether they also observed the change or not. The behavioral change can be the result of a firmware update or of an attack. To decide if the new behavior is legitimate, Sentinels rely on the majority's observation: If the change has been observed by the majority of Sentinels, it is considered as legitimate and will be added to the whitelist. Otherwise it will be considered as anomalous and will be blocked. This logic is based on the idea that if the majority of devices of the same type share the same behavior, it should be their "intended" behavior. Note that the "intended" behavior may be anomalous, however in this case the device should be considered itself untrusted and actions should be taken against the manufacturer. For example, in January 2020, Google revoked Xiaomi's access to the Google Home Hub ecosystem after users were able to view strangers' security cameras video feed~\cite{bbc_xiaomi_2020}.

        \subsubsection{Transparency \& auditing }
            \label{sec:serenity:behavior_change:auditing}
            In addition to Sentinels network filtering capabilities, the open and public nature of \name's blockchain introduces a new data source for cyber security experts, allowing them to follow and audit in real time the behavioral evolutions of IoT devices. This can be used to monitor growing threats, updates adoption rates, etc. For example, it is possible to watch the spread of a growing botnet by watching the rejected forks. Transparency and privacy concerns are discussed in Section~\ref{sec:discussion:privacy}.
     
    \subsection{Device onboarding}
        \label{sec:serenity:startpoint}
        When a new device is connected, Sentinels first go through a profiling phase to fingerprint the device before enforcing network filtering. This process is detailed below: 
        \begin{figure}[h!]
            \includegraphics[width=\linewidth]{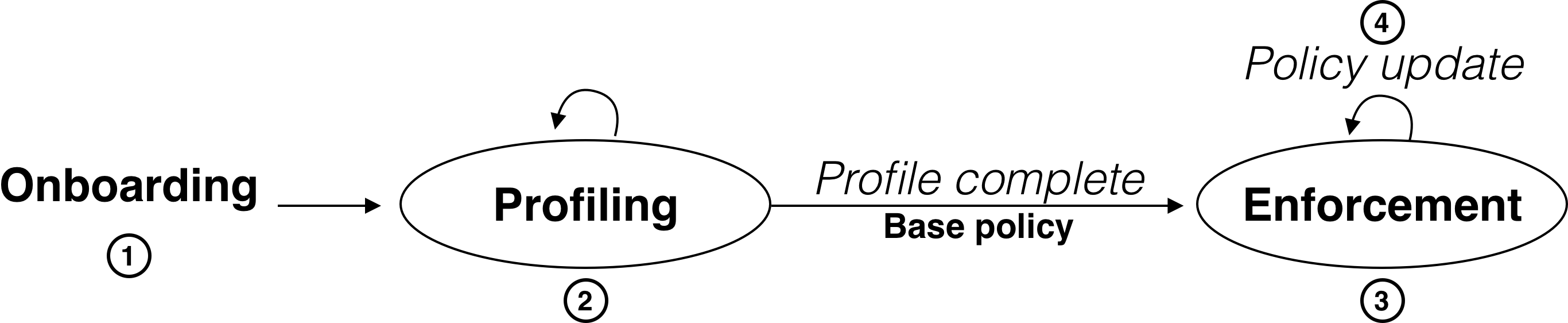}
            \caption{Device onboarding flow-chart.}
            \label{fig:startpoint-flow-chart}
        \end{figure}
        \begin{enumerate}
            \item \textbf{Onboarding:} The user buys a new device and connects it to the Sentinel. To connect the device to the Sentinel, the user uses the dedicated WiFi network broadcasted by the Sentinel.
            \item \textbf{Profiling:} During the profiling phase, the Sentinel allows all network traffic to and from the device and computes packet signatures for all the network connections to characterize the device. The connection signatures are used to register the new device to the blockchain regrouping all the devices with a similar network footprint. This profiling phase usually requires about 1 minute for most IoT devices tested during our evaluation (see Section \ref{sec:evaluation:virt}). If the corresponding blockchain doesn't exist (i.e. it is the first device with this network footprint to be connected to a Sentinel), a new device specific chain is initialized. 
            \item \textbf{Enforcement:} Once the device has been registered to a blockchain, the Sentinel downloads the blockchain to build the policy and begins blocking all the packets whose signatures don't match the policy. The Sentinel also starts reporting newly recorded packet signatures to the whitelist.
            \item \textbf{Behavior change / Policy update:} If the behavior of the protected devices changes, the Sentinel will vote to decide whether these changes need to be incorporated into the policy based on the majority observations. If only one Sentinel is registered to a blockchain, it can decide which packet signature to add to the policy. This can be problematic if the first Sentinel registered to a blockchain is malicious as it may incorporate anomalous packet signatures into the whitelist. This is discussed in detail in the security evaluation Section \ref{sec:evaluation:security}
        \end{enumerate}

\section{Evaluation}
    \label{sec:evaluation}
    To evaluate our system, we have carried out small scale experiments on real IoT devices to observe the system behavior in real-world use cases. We have also tested the scalability and robustness of \name during large scale simulations on Amazon AWS.

    \subsection{Implementation}
        \label{sec:evaluation:implementation}
        The proof of concept of the \name Sentinel is developed in node.js\footnote{https://nodejs.org/}, a cross platform, open source javascript runtime environment designed to build event-driven and asynchronous web applications. The blockchain component has been implemented from scratch. Sentinels communicate with peers using WebRTC\footnote{https://webrtc.org/} and websockets\footnote{https://developer.mozilla.org/docs/Web/API/WebSockets\_API}.
        In addition, we have developed a Web UI using VueJs framework\footnote{https://vuejs.org/} to monitor Sentinels in real time and help us run our experiments. Screenshots of the Web UI can be seen in Appendix~\ref{sec:appendix:screenshots}.   

    \subsection{Experiments on virtualized IoT devices and sentinels}
        \label{sec:evaluation:virt}
\begin{figure}[h!]
            \includegraphics[width=\linewidth]{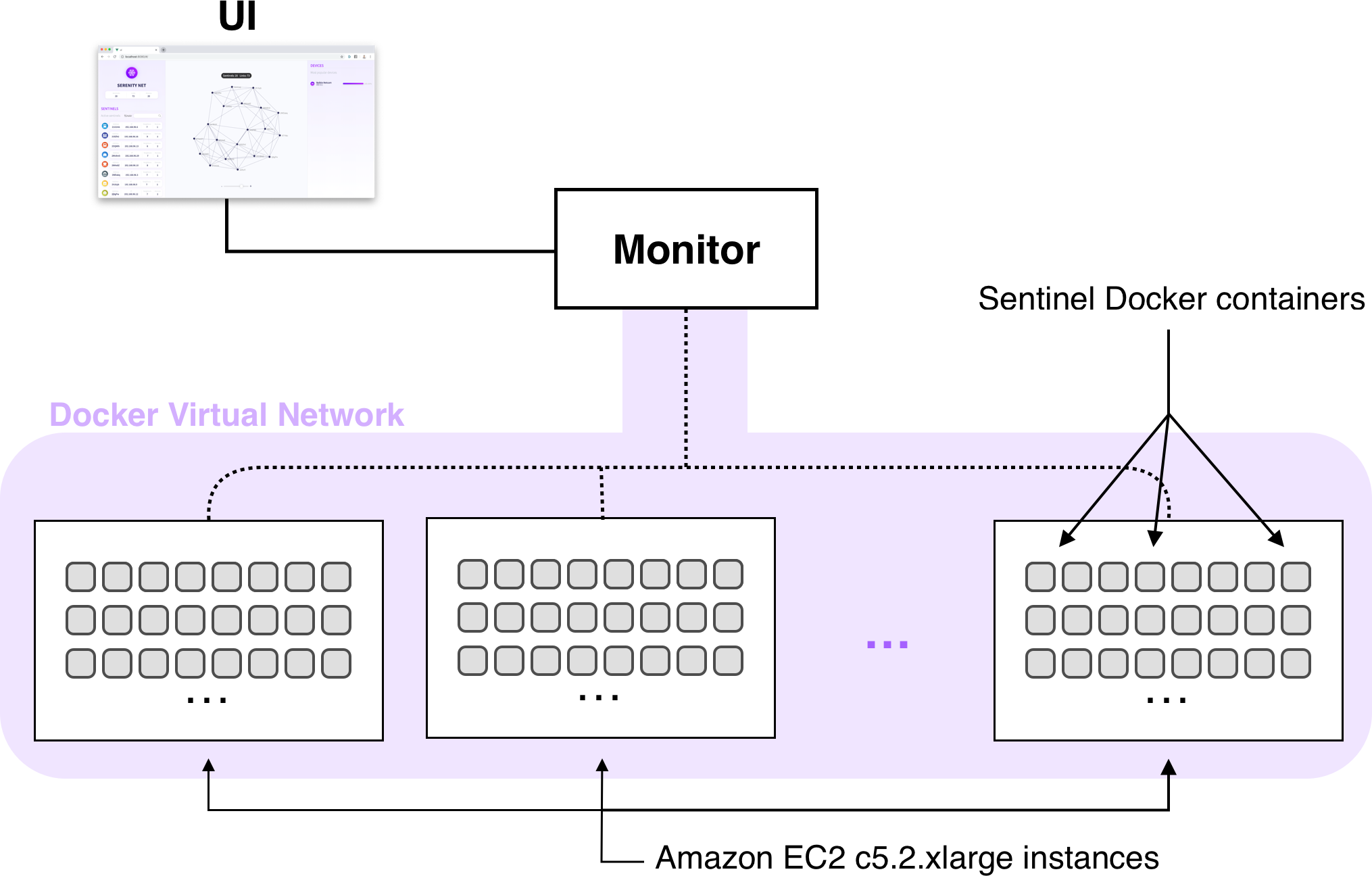}
            \caption{Simulation architecture of \name on Amazon AWS.}
            \label{fig:aws_architecture}
        \end{figure}

        To evaluate the Sentinels' behavior with different types of IoT devices and verify that our system can reach consensus and converge on a list of legitimate actions for the protected devices, we have set-up a virtualized testbed with 1000 Sentinels. Each Sentinel was run in a separated Docker container and we used Docker Swarm to orchestrate our cluster and deploy the Sentinels' containers on AWS instances. We used 10 Amazon EC2 c5.2.xlarge instances hosting each 100 Sentinels and connected through a Docker virtual network. For the simulations, Sentinels were running a simulated proof of work consensus algorithm in order to be able to simulate multiple Sentinels on a same host. We also developed a custom web app to monitor the state of each Sentinel to obtain a real time visualization of the Sentinel network, blockchains and evolution of policies. Figure \ref{fig:aws_architecture} shows the Amazon AWS architecture used for our evaluation.

        Each Sentinel was simulating a set of devices from the dataset of Alrawi et al.~\cite{alrawi_sok:_2019}. This dataset provides packet captures of an IoT network with 53 different devices for 9 continuous days. Based on these captures, we extracted the devices' behavior by isolating packets generating unique signatures to get a set of packets representative of each device's packet signatures. Devices were simulated by replaying these packets in random order and at random intervals to imitate the unpredictable aspect of user interactions (for a example a smart light bulb might be powered off during a period of time and the user can interact with it at any time).

        \begin{figure*}[h!]
            \includegraphics[width=\linewidth]{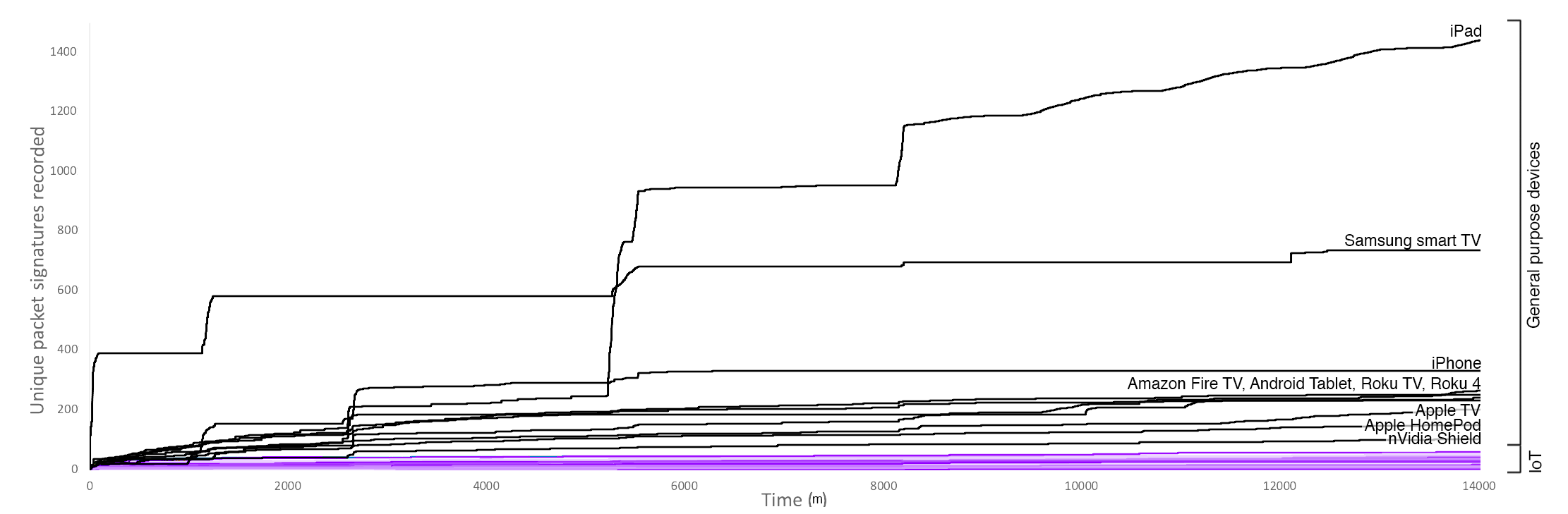}
            \caption{CDF of distinct packet signatures per device recorded over a 9-day period. Labeled lines identify general purpose devices.}
            \label{fig:evaluation:cdf}
        \end{figure*}

        \begin{figure*}[h!]
            \includegraphics[width=\linewidth]{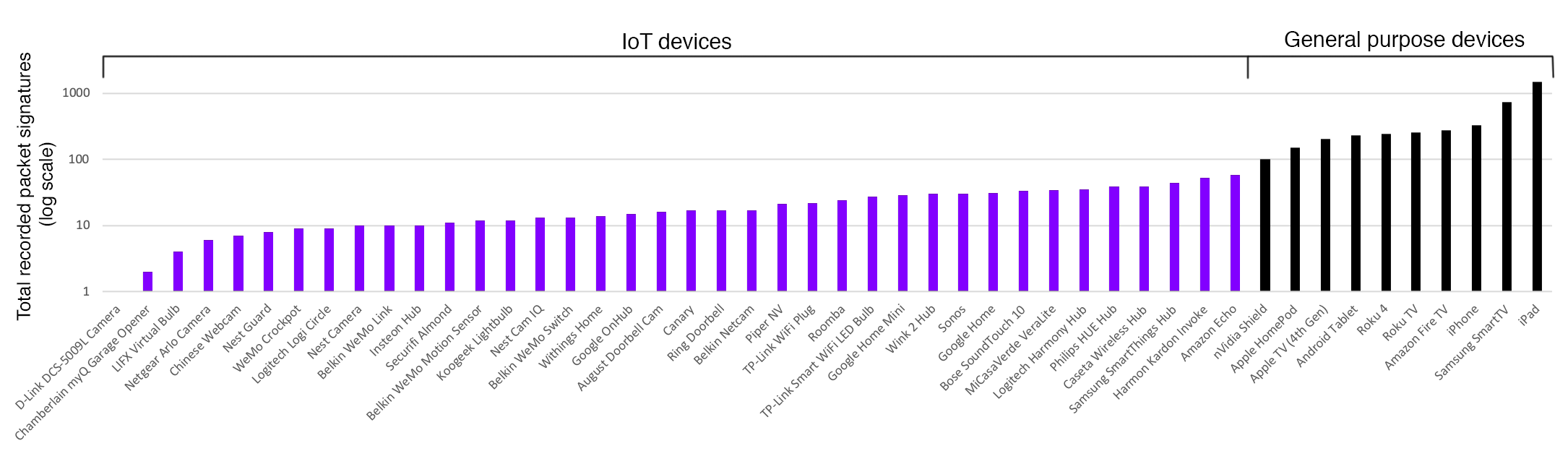}
            \caption{Total recorded packet signatures per device.}
            \label{fig:evaluation:fingerprints}
        \end{figure*}

        As shown in Figure \ref{fig:evaluation:cdf}, we clearly denote two classes of IoT devices: devices with a simple behavior characterized by a small number of packet signatures as Lifx Smart Bulb, TPLink WiFi plug or Nest Guard and other more complex multipurpose devices such as iPads, smart TVs, etc. 

        We also observe on Figure \ref{fig:evaluation:cdf} that devices with a simple functionality are characterized by a stable behavior overtime that doesn't change often. This validates our initial hypothesis that IoT devices' typical behavior only contains a small set of actions which remain constant over time. 
        
        During the experiments, we monitored the system and the devices' specific blockchains for different time periods (from 1 hour to multiple days) and with different number of Sentinels (from 20 to 1000); as expected, Sentinels were unable to converge on a list of legitimate packet signatures for general purpose devices as iPads, iPhones and Android tablets. However, for simpler devices Sentinels were able to converge and produced lists of trusted packet signatures. 
        
\begin{figure}
            \includegraphics[width=\linewidth]{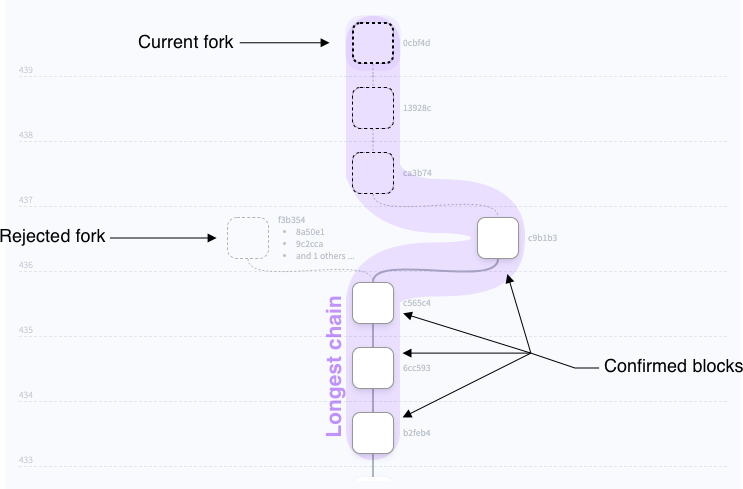}
            \caption{Rejected fork with anomalous packet signatures. The whitelist contains only packet signatures listed in the longest chain.}
            \label{fig:rejected_fork}
        \end{figure}

        We then subjected the lists of trusted packet signatures to attack data from the IoT network intrusion dataset of Hyunjae et al.~\cite{hyunjae_dataset_2019}\footnote{Based on this attack data, we simulated Mirai-style flooding and scan attacks. We manipulated the packets to reflect our IoT devices.}. To do so, we first injected malicious packets in a small number of Sentinels' devices' behavior. As expected, Sentinels monitoring "infected" devices successfully identified and blocked the malicious packets while keeping the device functional. Figure \ref{fig:rejected_fork} shows the identification and rejection of a fork with anomalous packet signatures. 

        We then attempted to determine the percentage of Sentinels monitoring compromised devices required to incorporate a new packet signature into the whitelist. This number characterises the breaking point where behavior changes are incorporated in the trusted list of packet signatures.
        
        During our experiments we observed that this breaking point occurs at around 51\%, meaning that 51\% of the Sentinels need to record a same packet signature to be authoritative on the longest chain and include the packet signature in the whitelist for a given device. Thus, popular devices are less likely to be attacked as more Sentinels need to be infected to incorporate malicious packet signatures in the whitelist. However, they also require more time for updates to be deployed as updates need to reach a greater number of devices before being trusted. 
        
        \subsection{Experiments on real IoT devices and sentinels}
            \label{sec:evaluation:real}
            We tested \name on real IoT devices to demonstrate that Sentinels can be easily deployed on a IoT network with minimal topology change and that our proof of concept also fits to real use. To do so, we installed the Sentinel software on 3 Raspberry PI 3 B+ configured as Wifi hotspots and used netfilterqueue\footnote{netfilterqueue is a wrapper around libnetfilter\_queue that gives access to the packets matched by specific iptables rules. More information: https://netfilter.org/projects/libnetfilter\_queue/} to intercept, inspect and block network packets forwarded by the Sentinels. We then connected 1 LIFX Mini Smart bulb to each Sentinel.
        
            \begin{figure}[h!]
                \includegraphics[width=\linewidth]{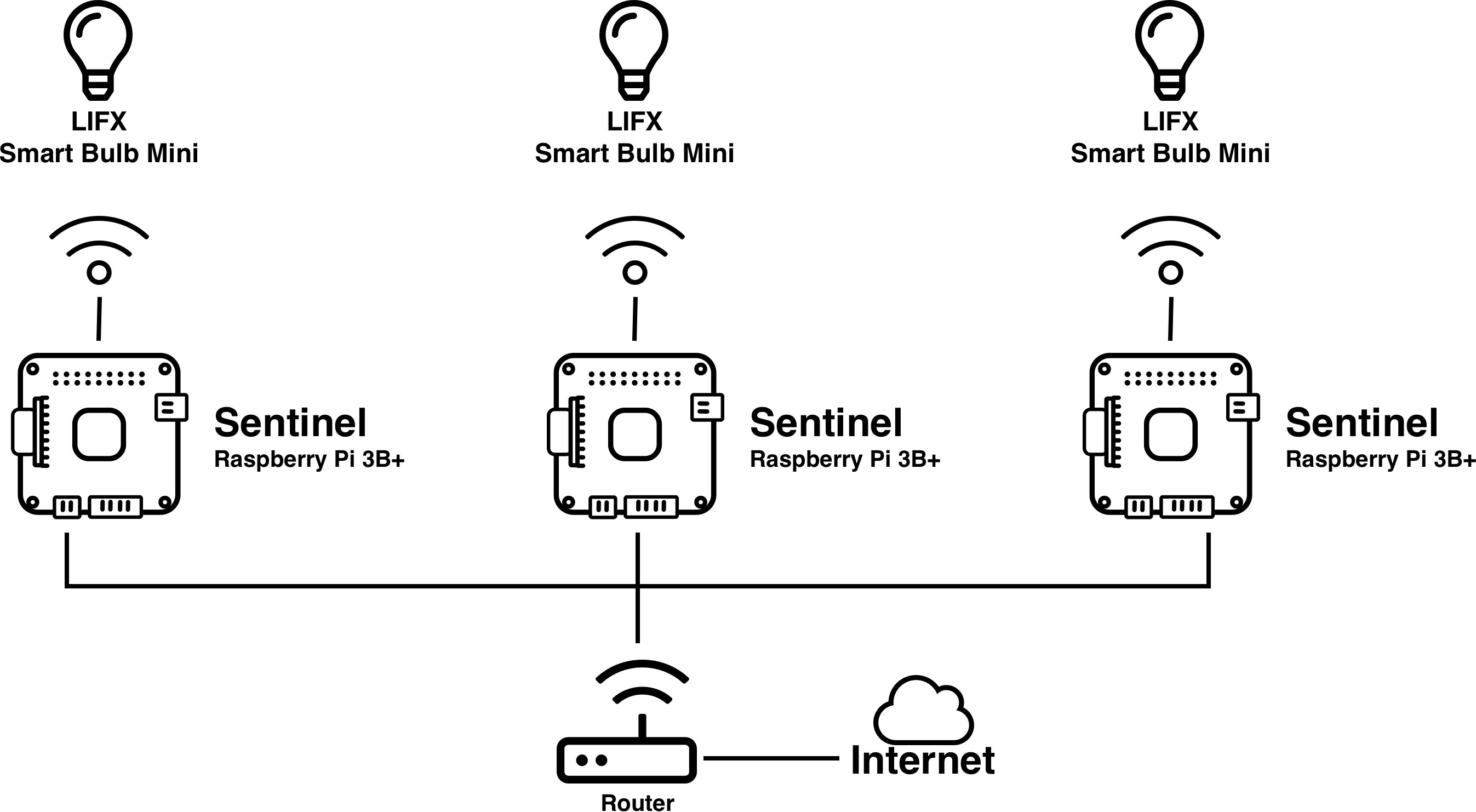}
                \caption{Physical devices used for our real experimental set-up.}
                \label{fig:evaluation:convergence:schema}
            \end{figure}
        
            This experiment was carried during a one-hour period during which we interacted with the devices through the manufacturers' mobile app. The 3 Sentinels were successfully able to record the packets from the bulbs and to converge on the list of the resulting packet signatures showing that our proof of concept also works with real life IoT devices and Sentinels. Moreover, during this experiment, there was no perceptible delay when interacting with the bulb through the LIFX mobile app and using \name.

            \begin{table}[h!]
                \centering
                \small
                \begin{tabular}{|l|l|l|}
                    \hline
                    \multicolumn{1}{|c|}{\textbf{Recorded packet}} & \multicolumn{1}{c|}{\textbf{Pkt. Signature}} & \multicolumn{1}{c|}{\textbf{Desc.}} \\ \hline
                    UDP time1.google.com R123 & 0cca40...aed4d4 & NTP \\ \hline
                    TCP 104.198.46.246 R56700 & 4e2b3d...2a4474 & LIFX API \\ \hline
                \end{tabular}
                \caption{Packet signatures recorded for the LIFX Smart Bulb during our experiment. Signatures have been truncated.}
            \end{table}
    
        \subsection{Blockchain performance evaluation}
            \label{sec:evaluation:blockchain}
            We have evaluated Sentinel performance and blockchain growth over a 24-hour experiment with 20 simulated Sentinels. The measurements show that the block sizes for the control and device chains tends to be constant over time. We estimated the size of the blockchains after one year of continuous operation of experimental proof of concept. Note that this code has not yet been optimized for wide-scale deployment, and many further improvements are possible, as discussed below. The control chain size depends on the number of distinct IoT device types protected by the Sentinels. We estimate the chain to grow to 511MB per year for 1 IoT device, and 511GB per year for 10,000 distinct IoT device types. We also observed that per-device chain size is consistent across all device types and should require around 474MB per year. Sentinels only need to download and maintain device chains corresponding to the device types they protect. Thus, a Sentinel with 10 different IoT devices would need to maintain a single copy of the control chain and 10 device chains. We note that many blocks in device chains will be empty and old blocks may contain signatures for packets that are no longer seen. There is thus an opportunity to periodically truncate old blocks containing no useful information and force Sentinels to submit fresh observations. This feature would prevent the blockchains to grow infinitely while allowing old policies to be updated by deleting outdated behaviors no longer in use by the majority of the devices.

            Our Sentinel performance measurements show that Sentinels use approximately 140MiB of RAM after running for 24 hours and that their network usage is correlated with blockchain growth. Network utilization varies between two Sentinels based on the number of distinct IoT devices they are protecting. Finally, Sentinels were using a simulated proof of work consensus algorithm to run the experiment. Their CPU usage is thus not representative of the usage one would observe if the Sentinels were using real proof of work instead. In the case of real proof of work, we expect CPU usage to be maxed out at 100\% for all Sentinels. Full measurements details can be found in Appendix~\ref{sec:appendix:blockchain_evaluation}

        \subsection{Security evaluation}
            \label{sec:evaluation:security}
            When a new IoT device is connected to a Sentinel, the Sentinel determines the whitelist to register based on the device behavior. Thus, devices behaving similarly will be grouped on the same whitelist and already compromised devices behaving differently will be assigned to a separate whitelist.
            Thanks to this mechanism all devices assigned to a same whitelist initially behave similarly. Uncompromised devices are thereby grouped together, and the corresponding whitelist will only contain legitimate packet signatures reflecting the behavior of these devices. In this section, we evaluate the different attacks vectors leading to successful attacks incorporating malicious packet signatures into the whitelists or exploiting devices to change their behaviors.
            \paragraph{Attacks against IoT during the profiling phase} During the profiling phase (see Section \ref{sec:serenity:startpoint}), Sentinels allow all the traffic and don't enforce any network filtering for the newly connected IoT device. Even if this phase only lasts a few minutes, an attacker could benefit from this short period of time to perform an attack. In this case, the attack will modify the network footprint of the device which will likely cause it to be registered to a different chain than other benign devices of the same type. This chain will regroup all the devices of this type that have produced the same network footprint during the profiling phase (i.e. all the devices of the same type that have been targeted by the same attack during the profiling phase) and the Sentinels will not filter the resulting malicious network connections.
            
            Devices could also already be infected when connected to Sentinels. In this case, if the infected devices' behavior is similar to benign devices of the same type, they will be registered to the same device chain as other devices and their malicious behavior will be filtered and blocked. However, if the infected devices' behavior is different, they will be registered to a chain with all the similarly infected devices as in the case of an attack during the profiling phase.
    
            \paragraph{Attacks against the IoT devices} Once the Sentinel is in the enforcement phase (see Section \ref{sec:serenity:startpoint}), we differentiate two types of attacks against the IoT devices. Attacks from the local network and attacks from the internet.
            
            Attacks from the local network will eventually succeed and compromise devices as Sentinels don't enforce any network filtering on the local network. However, Sentinels will block any behavior deviating from the specification trying to reach the internet. This protect against compromised local IoT devices trying to attack targets on the internet and restrict them to the strict behavior listed in the specifications. Moreover, thanks to the open nature of its blockchain, \name also provides a new kind of metrics allowing experts to monitor in real time the behavior of a large number of IoT devices. Thanks to this feature, \name provides transparency and can help cybersecurity experts to take actions and block threats as they grow.
            
            Attacks originating from a remote attacker targeting a specific IoT model will be blocked by Sentinels as long as only a minority of Sentinels observe the same attack pattern. Thus, these attacks grow in difficulty with more Sentinels protecting more devices of a specific model. For an attack to succeed, its network footprint has to be similar on a majority of Sentinels (meaning that Sentinels should record packets signatures with the same IP address). The attacker also needs to target 51\% of all devices in less than one block-interval (the time interval between blocks). For popular IoT devices, attacker unlikely have the resources to initiate an attack targeting simultaneously a large number of devices from a single host. Commonly, attackers rely botnets to carry massive attacks against IoT devices. IoT botnets such as Mirai\cite{antonakakis_understanding_nodate}, Brickerbot\cite{noauthor_brickerbot_nodate} and Hajime\cite{herwig_measurement_2019} share similar network footprints during the infection phase: they scan for listening telnet, ssh and http services and try to bruteforce the credentials. However, the source IP addresses of the attacker vary as botnets use infected devices to spread and infect new hosts. The packet signatures of each botnet attack will thus likely vary from one target to another as the IP and port used by the attacking device will vary. By design, \name should effectively block P2P botnets. Particularly those having multiple attackers. As these signatures are not consistent across all Sentinels, they are unlikely to be added to the whitelist and the attack will be blocked. 
            
            \paragraph{Attacks on the Sentinels} Sentinels act as both blockchain nodes and network traffic enforcement points. Thus, if a Sentinel is compromised, the attacker may insert or remove traffic rules arbitrarily\footnote{This is not unlike the security of a firewall or router, which should typically be better protected than internal hosts on the network.}. However, attacking a specific node does not allow the inclusion of malicious packet signatures into the blockchain, as these must still be validated and confirmed by the majority of the nodes. \name's blockchains are vulnerable to two types of majority attacks.
            \begin{itemize}
                \item Majority attacks against the control chain can succeed if an attacker has the computational capacity of more than half of all Sentinels. This adversary can win every proof-of-work round, allowing the inclusion of arbitrary packet signatures into any device blockchain/whitelist. This type of attack is devastating to the network since the whitelists are shared amongst all Sentinels. We discuss alternative consensus protocols to lower the probability of such an attack in Section \ref{sec:discussion:consensus}.
                \item Majority attacks against device chains can succeed if an attacker controls more Sentinels than half of all Sentinels registered on a specific device chain. This adversary will be able to append blocks to these chains more often than legitimate Sentinels which can result in incorporating malicious blocks into the longest device chain. This attack grows in difficulty with more Sentinels protecting the same device of a specific model and contributing to its chain. For popular IoT devices, with a large number of Sentinels registered on their chain, this attack's difficulty is similar to a majority attack on the control chain.
            \end{itemize}

\section{Limitations}
    \label{sec:limitations}
        Our system currently only monitors LAN-to-WAN connections so it does not protect IoT devices from other infected devices on the local network. While this limitation allows P2P infection methods to succeed on the local network, infected IoT devices will be unable to attack remote hosts; Sentinels will block the outgoing traffic that does not match whitelisted connections. 
        
        Our system is tailored to support IoT devices with a small network footprint. It is unclear, however, how effective the system can be in protecting IoT devices with richer network behavior (i.e., requiring a larger whitelist). An open question is whether any collaborative intrusion detection system (ours included) can converge on a set of connections that should be permitted. One strategy for complex devices is to make the packet signature algorithm less specific, but this may have the disadvantage of missing certain attacks. 

        In its current design, \name will block any user-defined connection to remote servers (e.g., a cloud-based FTP server for video stream backups). Connecting to user-defined endpoints will typically be blocked because connections to these arbitrary servers will not be observed amongst the broader population of Sentinels. One option to permit user-defined servers to be allowed is to enable a manual override in the user interface, allowing advanced users to allow specific connections without impacting the collective whitelisting feature. We will explore adding this feature in future work. 
        
        Finally, like any collaborative system, performance (both accuracy and resilience to attack) improves with the deployment of each additional node. By deploying more Sentinels, manipulating the whitelists (i.e., the blockchain) requires more effort from an attacker. Similarly, in the case of a small number of Sentinels, the likelihood of an attack influencing the packet signatures that get added to the whitelist is higher. 
        
\section{Discussion}
    \label{sec:discussion}

     \subsection{Consensus}
        \label{sec:discussion:consensus}
        In our proof of concept, we used Proof of Work for its implementation simplicity and wide availability. However, we expect that most Sentinels will be deployed on devices with limited computational power such as routers. As proof of work relies on computationally intensive problems to secure the chain, an attacker with large computing resources can easily overpower even a large number of routers. To mitigate this, an alternative consensus algorithm could be used. In addition to the consensus algorithms mentioned in Section~\ref{sec:background:blockchain}, a promising alternative is the \textit{proof of elapsed time} (PoET)~\cite{noauthor_poet_nodate}. This algorithm leverages Intel's Safe Guard Extension (SGX)~\cite{admin_intel_nodate} to execute the consensus algorithm in a verifiable environment. SGX allows to produce a certificate to attest that it's correctly running a given trusted code. PoET has the same objective as proof of work: to randomly delay block production so that it is evenly spread across the network over time. This is done by requiring nodes to run trusted code (certified through Intel's SGX). Unlike PoW, PoET doesn't require heavy computational effort and thus is less energy consumptive. However, this method also has limitations: It is tied to a given chip vendor so it would and relies entirely on a third party platform to work exposing it to vulnerabilities~\cite{217543, gotzfried_cache_2017,weichbrodt_asyncshock:_2016}.
    
    \subsection{Transparency and privacy}
        \label{sec:discussion:privacy}
        Blockchain systems, in particular those which are public tend to elicit privacy concerns since the ledger is replicated across all participants. By using \name, IoT device behaviour is published into an immutable public data structure that can help users understand the expected functionality of a device before it is purchased. Auditors and regulators can use this information as well to inform regulation of future devices. The disadvantage of this transparency is that directly connected nodes can query each other for the availability of blocks corresponding to particular chain. This could allow an attacker who knows the mapping between a device and a chain to ask whether a Sentinel has that device. This attack could be mitigated by throttling the number of requests allowed per source IP address and by adding an exponentially increasing delay for every subsequent request from the same source. 
        
    \subsection{Usability}
        \label{sec:discussion:usability}
        Despite its complexity, \name can fully operate without any user interaction to protect the vast majority of simple IoT devices. We expect that this zero-configuration will encourage adoption even by non-expert users. However, in the event of software updates that change the behaviour of an IoT device, the system will prevent new functionality from working until more than majority of Sentinels observe the same behaviour on their devices. The research community has not yet measured the speed of deployment of software updates on IoT. Lack of updating may leave early adopters without the ability to use the new features. One might argue that for security reasons, waiting for the majority of devices to upgrade is safer, but some users may want the latest features as soon as possible. 
        
        In future, we may add a manual override to allow expert users to clear the currently learned behaviour of their device and treat it as new after the update has been applied which will likely force the device onto a different chain. 
        
        The system provides usability for common main stream users. It doesn't require any user interaction to work and will protect devices once they are connected to it. For real world deployment, \name could partner with ISPs to deploy the Sentinel software on home gateways.

\section{Conclusion}
    \label{sec:conclusion}
    IoT devices in smart homes are often unnecessarily overprivileged, increasing the risk of compromise and impact of attacks. This paper explored leveraging blockchain technology to assist in determining a strict specification of essential network behavior of IoT devices. We presented \name as a proof of concept network policy management and enforcement system that can operate with little to no user input. Our evaluation shows that the system is able to converge on small network security policies for many simple consumer IoT devices without requiring changes to firmware, software, or apps, and without requiring vendor buy-in. The consensus algorithm forces attackers to execute majority attacks to make changes to those policies. While implementation and deployment questions remain, we hope that \name can be viewed as a first step toward blockchain-based network security policy enforcement systems.

\section*{Acknowledgments}
    We thank José Fernandez and Jeremy Clark for helping shape initial versions of this work. We also thank Xavier de Carné de Carnavalet and Paul van Oorschot for their insightful feedback. This research is partially supported by the Natural Sciences and Engineering Research Council of Canada (NSERC Discovery Grant RGPIN-2018-04468).
 
\bibliographystyle{plain}
\bibliography{arxiv}

\appendix
\section{Appendices}
    \subsection{Blockchain evaluation}
        \label{sec:appendix:blockchain_evaluation}
        This section evaluates the capacity of our system to run on the long term. During our experiments we have measured the growth of the blockchains and monitored the runtime metrics of the Docker containers running the Sentinels.

        \paragraph{Blockchains size} Sentinels store blocks as JSON files. To measure the blockchain growth we connected to different Sentinels during a 24-hour experiment and recorded the number of stored blocks as well as the size of the blocks directory after 1 hour, 5 hours and 24 hours. For this experiment we used 20 simulated Sentinels with one Belkin Netcam connected. The number of Sentinels in the network does not influence the blockchain size as the block production rate is fixed and determined by the consensus algorithm. Sentinels also delete rejected forks blocks as soon as they converge on a longest chain.

        Table \ref{appendix:controlchain} shows that the control chain block size tends to be constant overtime. The block size for the control chain is determined by the number of different IoT devices types protected by the Sentinels. Indeed, each device type has is own device chain and each device chain is indexed into the control chain. Control chain blocks list the block headers of the latest produced blocks from the device chains, containing at maximum the number of device chains, block headers. Block headers are SHA256 hashes and have a fixed size of 32 bytes. It is thus easy to compute the control chain block size for a given number of IoT devices. Based on our measurements, the control chain with 1 IoT device should be around 511MB after one year running.
        \[1.4MB * 365 days = 511MB/year\]
        If we consider 10K different IoT device types protected by our Sentinels, the control chain should be around 511GB after one year running.
        \[4384 blocks * 10 000 * 32B + 1.4MB = 1.4GB/day\]
        \[1.4GB * 365 days = 511GB/year\]

        Table \ref{appendix:devicechain} shows that the device chain block size also tends to be constant. Indeed, most of the blocks in device chains are empty as blocks list packet signatures of newly observed behaviors. In the long run, the majority of blocks will thus be empty as new behaviors are rarely recorded. Based on our measurements,  device chains should be around 474MB after one year running.
        \[1.3MB * 365 days = 474MB/year\]

        Sentinels are required to maintain a copy of the control chain. However, they only require to download and maintain the devices chains corresponding to the device types they protect. Thus, a Sentinel with 10 different IoT devices would need to maintain a copy of the control chain and 10 device chains.

        Future work will explore a block expiration feature where outdated blocks will be deleted. This feature would prevent the blockchains to grow infinitely while allowing old policies to be updated by deleting outdated behaviors no longer in use by the majority of the devices. 

        \begin{table}[h!]
            \begin{tabular}{|l|l|l|l|}
            \hline
            \textbf{Elaps. Time} & \textbf{Nb. of blocks} & \textbf{Size} & \textbf{AVG block size} \\ \hline
            1 hour & 205 & 64KB & 312B \\ \hline
            5 hours & 918 & 291KB & 316B \\ \hline
            24 hours & 4384 & 1.4MB & 316B \\ \hline
            \end{tabular}
            \caption{Block size measurements for the control chain with 1 IoT device}
            \label{appendix:controlchain}
        \end{table}

        \begin{table}[h!]
            \begin{tabular}{|l|l|l|l|}
            \hline
            \textbf{Elaps. Time} & \textbf{Nb. of blocks} & \textbf{Size} & \textbf{AVG block size} \\ \hline
            1 hour & 193 & 59KB & 304B \\ \hline
            5 hours & 906 & 268KB & 296B \\ \hline
            24 hours & 4367 & 1.3MB & 295B \\ \hline
            \end{tabular}
            \caption{Block size measurements for a device chain}
            \label{appendix:devicechain}
        \end{table}

\begin{figure*}[h!]
            \includegraphics[width=\linewidth]{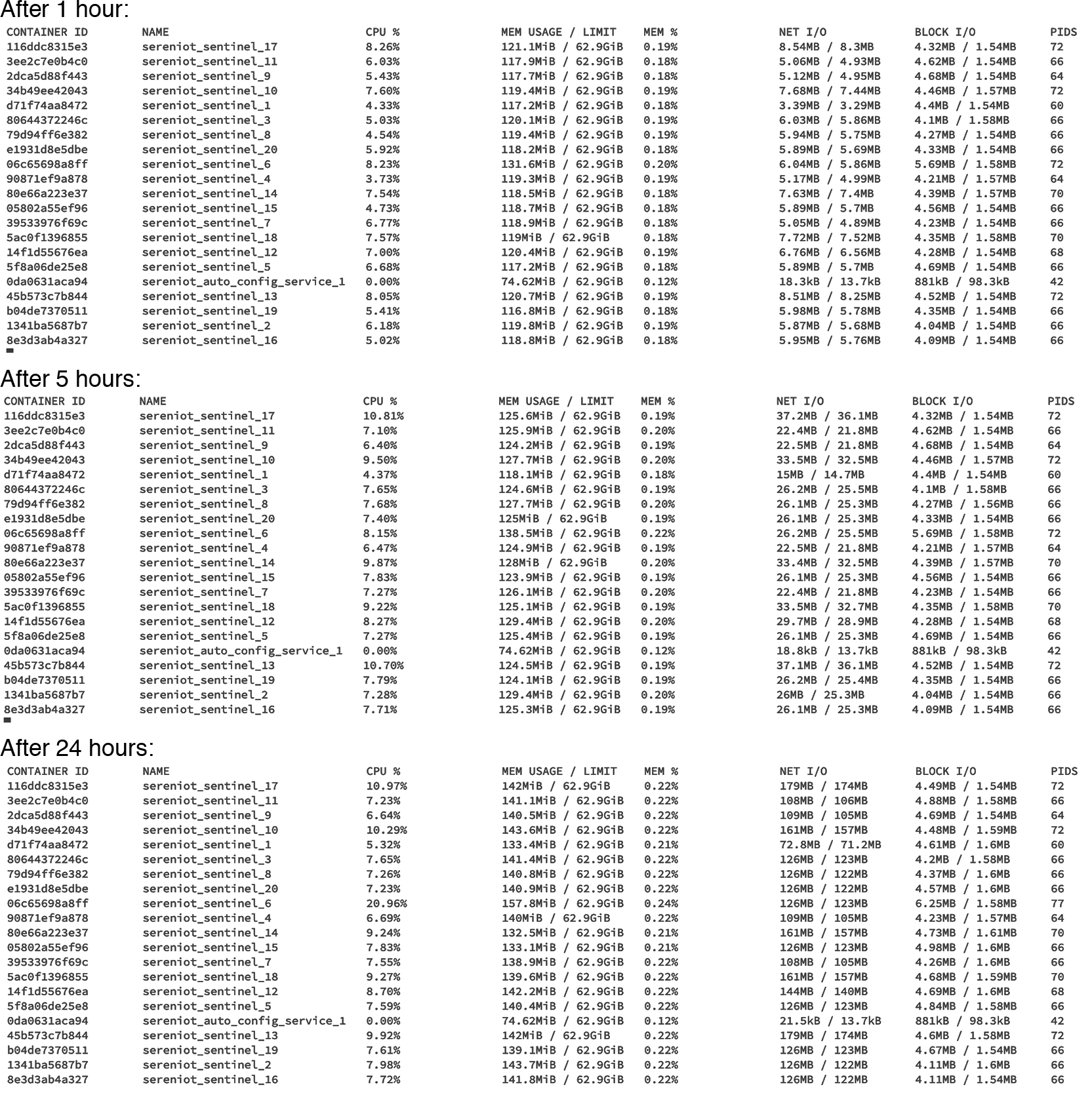}
            \caption{Screenshot of the Sentinels metrics.}
            \label{fig:screenshot:metrics}
        \end{figure*}

        \paragraph{Sentinels runtime metrics} To record Sentinels' metrics, we used the \textit{docker stats} command\footnote{More information on the Docker runtime metrics can be found here: https://docs.docker.com/config/containers/runmetrics/}. We recorded the metrics for 20 simulated Sentinels during a 24 hours experiment after 1 hour, 5 hours and 24 hours. These metrics show the CPU, memory and network usage. We observed that Sentinels use approximately 140MiB of RAM after running for 24 hours and that their network usage is correlated with the blockchains growth. Sentinels only download blocks for the control chain and for the device chains they are registered to. The network usage varies between two Sentinels based on the number of different IoT devices they are protecting. Finally, Sentinels were using a simulated proof of work consensus algorithm to run the experiment. Their CPU usage is thus not representative of the usage one would observe if the Sentinels were using real proof of work instead. In the case of real proof of work we expect CPU usage to be maxed out at 100\% for all Sentinels. Sentinels' metrics records can be found in Figure~\ref{fig:screenshot:metrics}.

    \subsection{Web UI Screenshots}
        \label{sec:appendix:screenshots}
\begin{figure*}[]
            \includegraphics[width=\linewidth]{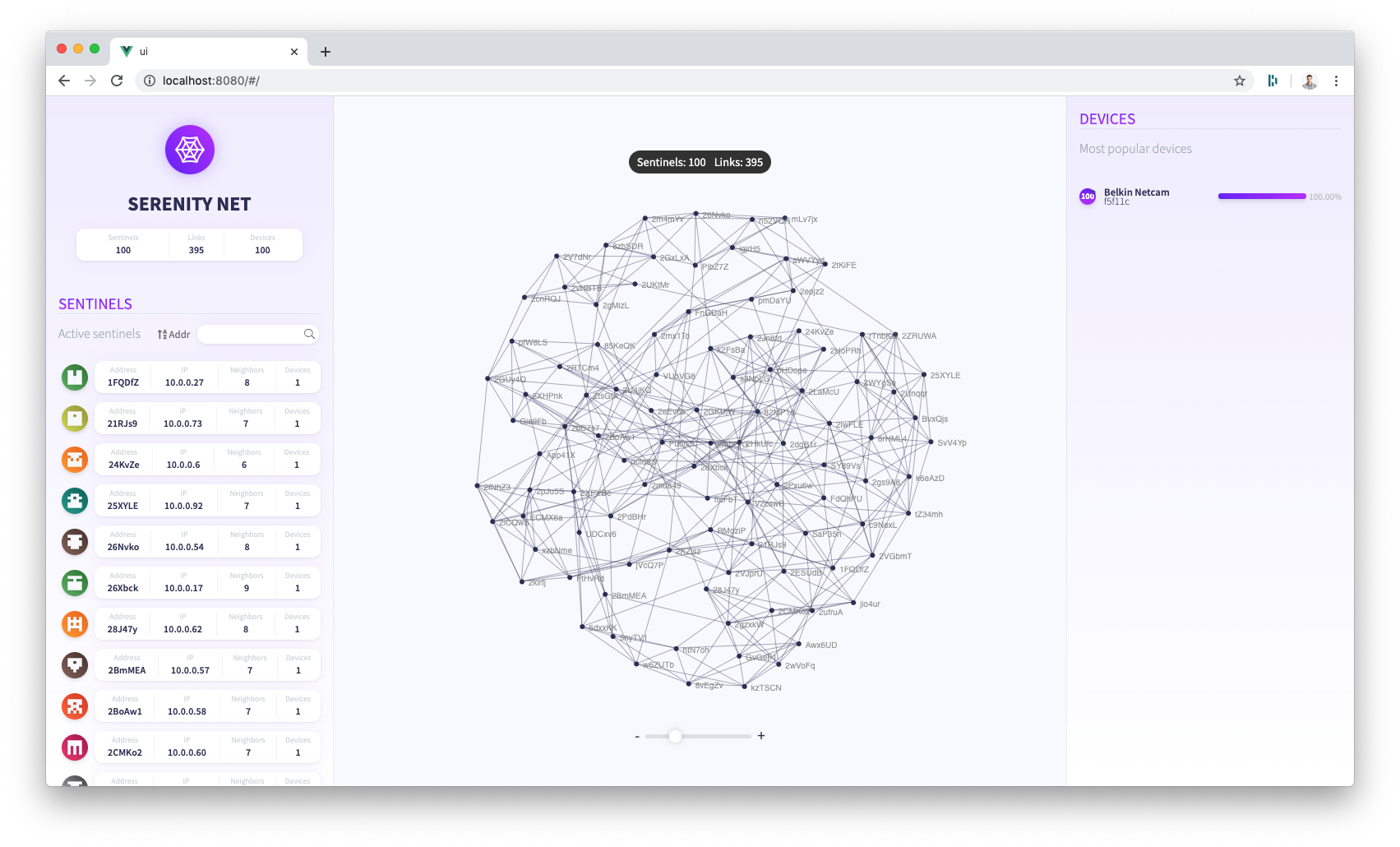}
            \caption{Screenshot of the Network view of the Web UI.}
            \label{fig:screenshot:ui_network}
            
            \includegraphics[width=\linewidth]{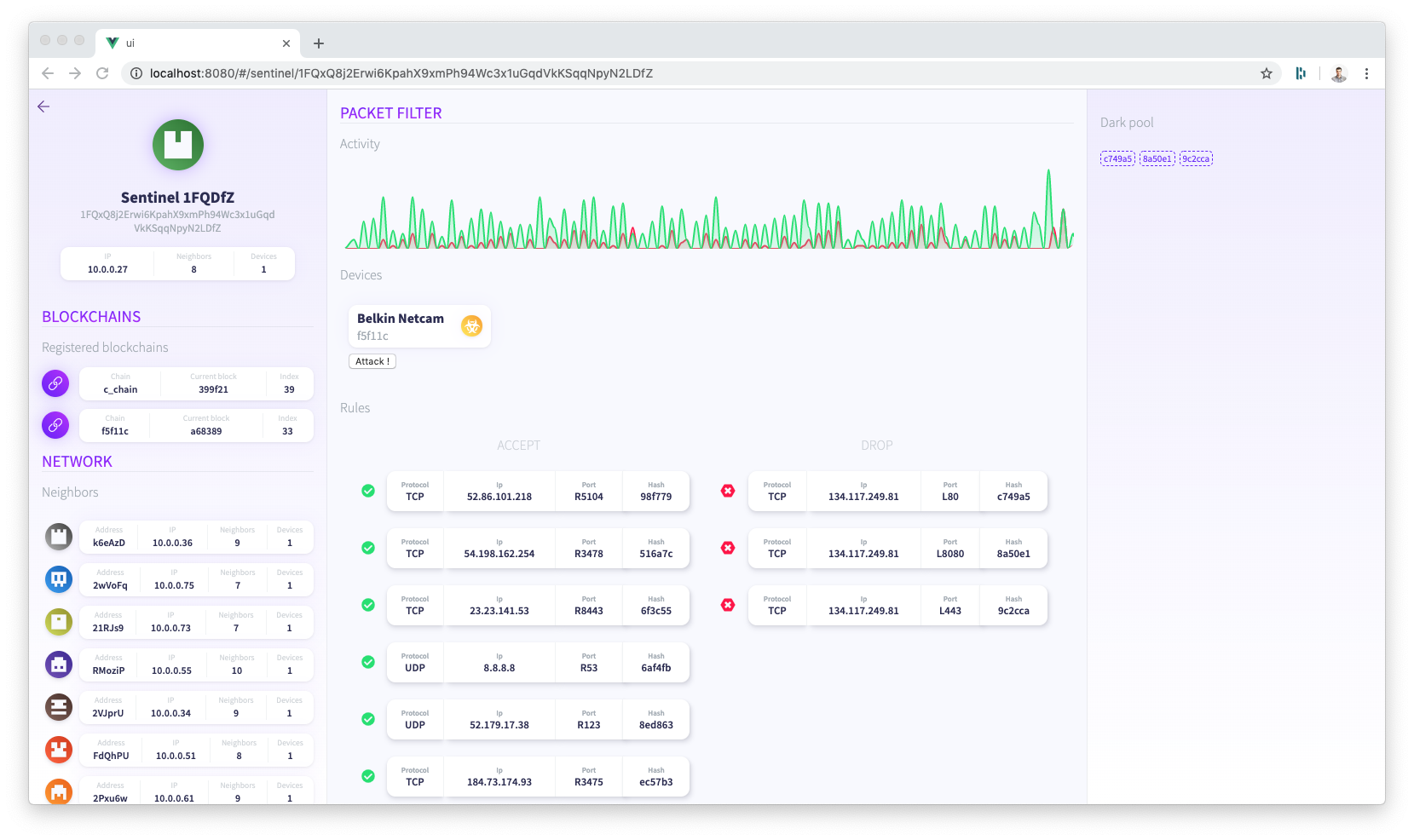}
            \caption{Screenshot of the Sentinel view of the Web UI.}
            \label{fig:screenshot:ui_sentinel}
        \end{figure*}

\begin{figure*}[]
            \includegraphics[width=\linewidth]{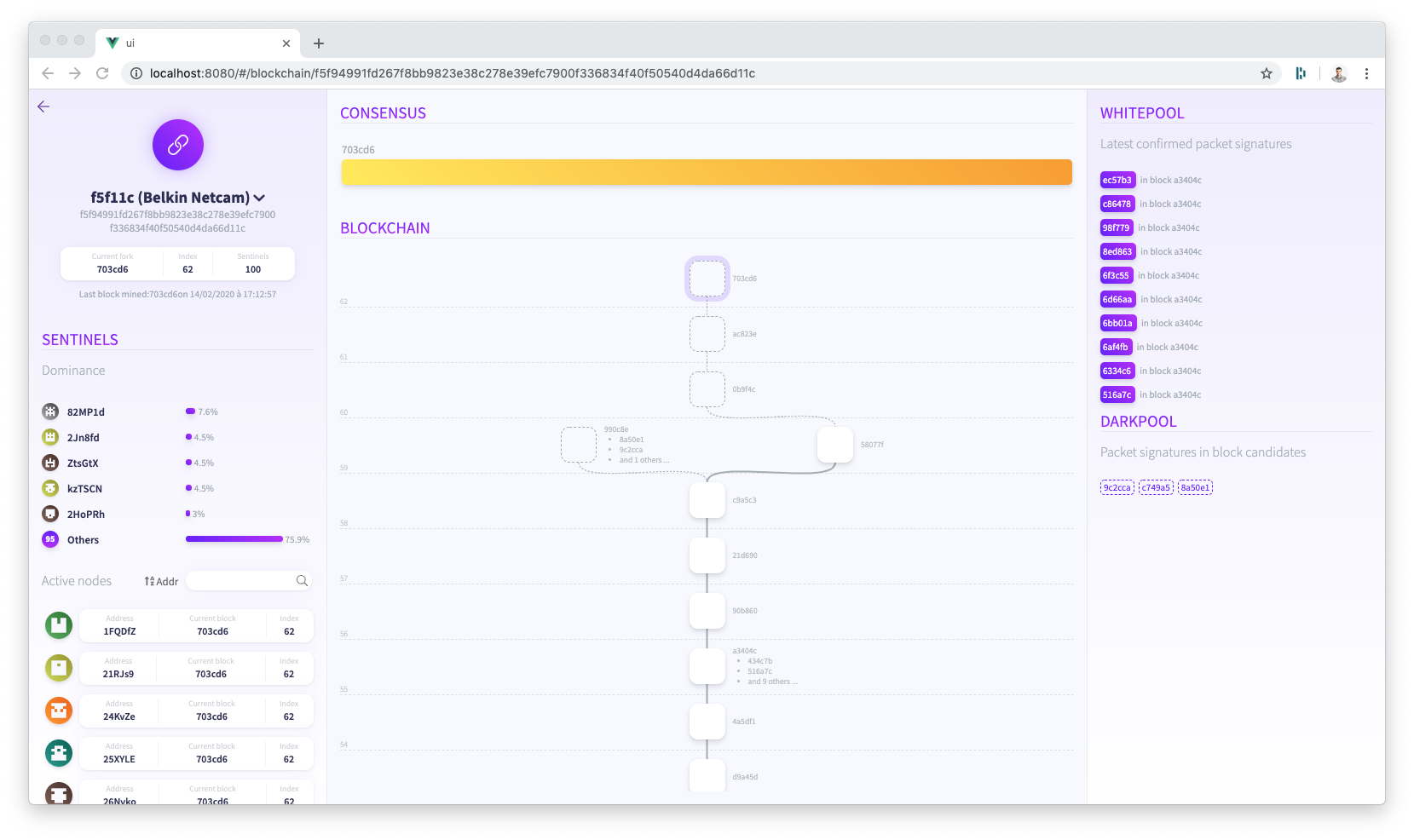}
            \caption{Screenshot of the Blockchain view of the Web UI.}
            \label{fig:screenshot:ui_blockchain}
        \end{figure*}

        Figures \ref{fig:screenshot:ui_network}, \ref{fig:screenshot:ui_sentinel} and \ref{fig:screenshot:ui_blockchain} show screenshots of the Web UI during a simulation with 100 Sentinels.

\end{document}